\documentclass[prl,superscriptaddress,floatfix,citeautoscript,longbibliography,twocolumn]{revtex4-2}
\usepackage{amsmath}
\usepackage{xcolor}
\usepackage{braket}
\usepackage{hyperref}
\usepackage{float}
\usepackage{graphicx} % Required for inserting images
\usepackage{hyperref}
\usepackage{float}
\usepackage{graphicx}
\usepackage{listingsutf8}
\usepackage{tcolorbox}
\usepackage{colortbl}
\usepackage{multirow}
\newcommand{\lcpq}{Laboratoire de Chimie et Physique Quantiques, CNRS, Universit\'e de Toulouse and European Theoretical Spectroscopy Facility, 118 Route de Narbonne, F-31062 Toulouse, France}
\newcommand{\lpt}{Laboratoire de Physique Th\'eorique, CNRS, Universit\'e de Toulouse and European Theoretical Spectroscopy Facility, 118 Route de Narbonne, F-31062 Toulouse, France}

\newcommand{\br}{\mathbf{r}}

\begin{document}

\title{Direct and inverse photoemission spectra from the screened multichannel Dyson equation}

\author{Pina Romaniello}
\affiliation{\lpt}
\email{pina.romaniello@irsamc.ups-tlse.fr}

\author{J.~Arjan Berger}
\affiliation{\lcpq}
\email{arjan.berger@irsamc.ups-tlse.fr}

\begin{abstract}
We present the screened multichannel Dyson equation for the simulation of both direct and inverse photoemission spectra from first principles.
The screened multichannel Dyson equation improves upon the standard multichannel Dyson equation by correctly including the screening of all particle-particle and electron-hole interactions due to the presence of the other electrons.
Using the example of bulk silicon, we demonstrate that the screened multichannel Dyson equation can capture the main features of the direct and inverse photoemission spectra.
In particular, it captures the correct position of the silicon plasmon satellite, unlike standard many-body approaches such as $GW$, which strongly overestimates the binding energy of this satellite.
Finally, we show that also the standard multichannel Dyson equation and the second-Born approximation strongly overestimate the binding energy of the plasmon satellite,
thus demonstrating the importance of properly screening all particle-particle and electron-hole interactions.
\end{abstract}

\maketitle
Direct and inverse photoemission spectroscopy probe the occupied and unoccupied states of materials, respectively, and thus provide valuable information about their electronic structure~\cite{hufner}.
Theoretical simulations of these spectra from first principles are crucial for analysing and understanding the experimental spectra as well as for predicting the electronic properties of novel materials.

For almost 40 years the $GW$ approximation~\cite{Hed65,Hybertsen_1986,Ary98,Hed99,Reining_2018,Golze_2019}, has been the method of choice to simulate direct and inverse photoemission spectroscopy.
While $GW$ has been very successful, in particular in describing band structures, it also has shown several shortcomings~\cite{Lan12,Ber14,Sta15,Loo18,Ver18,Tar17}.
One of the most important failures of $GW$ is its description of satellite structures in the spectral functions of molecules and materials.
In particular, the binding energies of satellites are largely overestimated by $GW$~\cite{Lan70,Ary96,Guz11,riva_prl}.
This is especially true when the electron correlation is strong.
However, this failure also appears in weakly correlated materials.
A well-known example is the plasmon satellite in the valence spectrum of silicon.
While in the experiment the satellite is centred at a binding energy of about 17 eV below the valence quasi-particles, corresponding to the bulk-plasmon frequency of silicon, 
in the $GW$ spectral function it appears at much lower energy, centred at a binding energy of about 25 eV below the valence quasi-particle energies.
We note that the correct position of the satellite can be obtained using the $GW$+cumulant approach which accurately describes plasmon-induced satellites~\cite{Ary96,Guz11,Lischner_2013,Zhou_2015,Gumhalter_2016} .

We recently introduced multichannel Dyson equations (MCDEs) in which two or more independent-particle many-body Green's functions are coupled through a multichannel self-energy~\cite{riva_prl,riva_prb,riva_prb_25,Paggi_JCP_2025}.
In particular, the (3,1)-MCDE simulates photoemission spectra by coupling a 1-body and 3-body independent-particle Green's function.
Unlike the standard single-channel Dyson equation, the (3,1)-MCDE naturally puts quasiparticles and satellites corresponding to electron-hole excitations on equal footing.
We have also proposed an approximation to the multi-channel self-energy and shown that it produces the exact spectral function of the Hubbard dimer at any interaction strength as well as accurate spectral functions for the extended Hubbard dimer\cite{riva_prl,Paggi_JCP_2025}.
This demonstrates the potential of the (3,1)-MCDE to simulate spectra of strongly correlated materials.

The current approximation to the multichannel self-energy contains all many-body contributions in both the particle-particle and electron-hole channels that are of first order in the bare Coulomb interaction.
However, as explained in detail in Ref.~\cite{riva_prb}, solving the (3,1)-MCDE yields infinite partial sums. Therefore, the (3,1)-MCDE implicitly includes contributions to all orders in the bare Coulomb interaction.
We have shown that by resumming these infinite partial sums, some of them correspond to screened Coulomb interactions~\cite{riva_prb}.
However, not all the Coulomb interactions in the multichannel self-energy are screened when solving the (3,1)-MCDE.
This might not be a problem when describing atoms and small molecules but could severely limit its application to large molecules and materials.
For these systems the screening of the particle-particle and electron-hole interactions by the electron cloud of the other electrons is crucial for a correct description of their electronic structure.

Therefore, in this work we present the screened multichannel Dyson equation which puts all interactions in the multichannel self-energy on equal footing
 by explicitly screening all Coulomb interactions that are not already naturally screened when solving the (3,1)-MCDE.
We achieve this by replacing the bare Coulomb potential in those interactions by a statically screened Coulomb potential.
We will show that this modification is crucial to correctly describe the plasmon satellite in silicon.
We note that the introduction of the screened potential has also been used to correctly describe other kinds of many-body excitations~\cite{Deilmann_2016,Torche_2019,Sab22,Dis23-1,Venkatareddy_2025}.

The (3,1)-MCDE~\cite{riva_prl,riva_prb} couples the independent-particle one-body Green's function $G^0_{1}(\omega)$ to the 2-hole-1-electron ($2h1e$) and 2-electron-1-hole ($2e1h$) channels of the independent-particle 3-body Green's function $G_{3}^{0,3\text p}(\omega)$~\cite{Riv22}. It is given by
\begin{equation}
\label{Eqn:MCDE}
    G_{3}(\omega)=G^0_{3}(\omega)+G^0_{3}(\omega) \Sigma_{3}(\omega)G_{3}(\omega),
\end{equation}
where $G^0_{3}(\omega)$ is defined as
\begin{equation}
    G^0_{3}(\omega)=\begin{pmatrix}
        G^0_{1}(\omega) & 0 \\
        0 & G_{3}^{0,3\text p}(\omega)
    \end{pmatrix}.
\end{equation}
It is convenient to express the (3,1)-MCDE in the basis of the independent-particle wave functions $\phi_i$ since $G^0_{3}(\omega)$ is diagonal in this basis.
Within this basis $G^0_{1}(\omega)$ and $G_{3}^{0,3\text p}(\omega)$ are given by
\begin{align}
    G^0_{i;m}(\omega)&=\frac{\delta_{im}}{\omega-\epsilon^0_i+i\eta\text{sign}(\epsilon^0_i-\mu)}, \\ 
    G_{i>jl;m>ok}^{0,3 \text p}(\omega)&=
    \frac{\delta_{im}\delta_{jo}\delta_{lk}(f_i-f_l)(f_j-f_l)}{\omega-\epsilon^0_i-(\epsilon^0_j-\epsilon^0_l)+i\eta\text{sign}(\epsilon^0_i-\mu)}.
\end{align}
in which $\epsilon^0_i$ and $f_i$ are the energy and occupation number corresponding to $\phi_i$.

In our approximation the multichannel self-energy $\Sigma_3$ is static and contains all interactions up to first order. It is given by
\begin{align}
    \Sigma_3=\begin{pmatrix}
        \Sigma^{1 \text p} & \Sigma^{\text 1p/3p}\\
        \Sigma^{\text 3p/1p} & \Sigma^{3\text p}
    \end{pmatrix}.
\end{align}
The explicit expressions of the contributions to $\Sigma_3$ depend on the independent-particle method employed to generate the single-particle wave functions and energies.
Here we will restrict ourselves to the Hartree-Fock method for which the various contributions are defined as
\begin{align}
    \Sigma^{1 \text p}_{i;m}&=0, \\
       \Sigma^{\text 1p/3p}_{i;mok}&= v_{ikom}-v_{ikmo}, \\
     \Sigma^{\text 3p/1p}_{ijl;m}&= v_{ijlm} - v_{ijml}, \\
      \label{Eqn:Sigmabody}
    \Sigma^{3 \text p}_{ijl;mok}&=  [(1 - f_i) (1 - f_j)f_l - f_if_j(1 - f_l)]
    \\ \nonumber &\times [\delta_{lk} (v_{ijom} - v_{ijmo})\nonumber + \delta_{mj}(v_{iklo} - v_{ikol}) 
    \\ \nonumber &+ \delta_{io} (v_{jklm} - v_{jkml}) - \delta_{oj} (v_{iklm}  - v_{ikml}) 
    \\ \nonumber &-  \delta_{im}  (v_{jklo} - v_{jkol})],
\end{align}
where the matrix elements of the Coulomb potential $v(\mathbf{r}_1,\mathbf{r}_2)=|\mathbf{r}_1 - \mathbf{r}_2|^{-1}$ are defined as
\begin{equation}
    v_{ikom}=\iint dx_1 dx_2 \phi^*_i(x_1)\phi^*_k(x_2)v(\mathbf{r}_1,\mathbf{r}_2)\phi_o(x_2)\phi_m(x_1).
\end{equation}
We note that the $(3,1)$-MCDE naturally combines particle-particle and electron-hole channels. Indeed the first term in Eq.~\eqref{Eqn:Sigmabody} is a particle-particle contribution while the last four terms are electron-hole contributions.

As explained in detail in Ref.~\cite{riva_prb}, the above multi-channel self-energy contains interactions that are naturally screened owing to the infinite partial sums generated by solving the (3,1)-MCDE for $G_3(\omega)$ in Eq.~\eqref{Eqn:MCDE}.
However, not all interactions in the multi-channel self-energy are screened leading to an imbalance between screened and unscreened interactions.
In order to put all interactions on equal footing, we introduce a new multichannel self-energy in which we screen all Coulomb interactions which are not naturally screened when solving the (3,1)-MCDE.
The screened multichannel self-energy $\Sigma_3^{scr}$ is thus defined as
\begin{align}
    \Sigma_3^{scr}=\begin{pmatrix}
        \Sigma^{1 \text p} & \Sigma^{\text 1p/3p}\\
        \Sigma^{\text 3p/1p} & \Sigma^{3\text p, scr}
    \end{pmatrix},
\end{align}
where $\Sigma^{1 \text p}$, $\Sigma^{\text 1p/3p}$, and $\Sigma^{\text 3p/1p}$ have been defined above and
\begin{align}
    \Sigma^{3 \text p, scr}_{ijl;mok}&=  [(1 - f_i) (1 - f_j)f_l - f_if_j(1 - f_l)]
    \\ \nonumber &\times [\delta_{lk} (W_{ijom} - W_{ijmo})\nonumber + \delta_{mj}(W_{iklo} - v_{ikol}) 
    \\ \nonumber &+ \delta_{io} (W_{jklm} - v_{jkml}) - \delta_{oj} (W_{iklm}  - v_{ikml}) 
    \\ \nonumber &-  \delta_{im}  (W_{jklo} - v_{jkol})].
\end{align}
The matrix elements of the screened Coulomb potential $W$ are given by
\begin{equation}\label{potential:eq}
    W_{ikom}=\iint dx_1 dx_2 \phi^*_i(x_1)\phi^*_k(x_2)W(\mathbf{r}_1,\mathbf{r}_2)\phi_o(x_2)\phi_m(x_1),
\end{equation}
where
\begin{equation}
W(\mathbf{r}_1,\mathbf{r}_2) = \int d\br_3 \epsilon^{-1}(\mathbf{r}_1,\mathbf{r}_3,\omega=0)  v(\mathbf{r}_3,\mathbf{r}_2),
\end{equation}
in which $\epsilon(\omega)$ is the dielectric function.

It is instructive to downfold the screened multichannel self-energy to the corresponding 1-body correlation self-energy $\Sigma^c_1(\omega)$.
The latter is given by 
\begin{align}
\Sigma^c_{i;\bar{m}}(\omega) &= \Sigma^{1p/3p}_{i;mok} G_{mok;rst}^{\text{3p}}(\omega)\Sigma^{3p/1p}_{rst;\bar{m}},
\label{Eqn:downfold}
\end{align}
where the summation over repeated indices is implied and
\begin{align}
G_{mok;rst}^{\text{3p}}(\omega) &=\left[G^{0,3 \text p,-1}_{\bar{m}\bar{o}\bar{k};\bar{r}\bar{s}\bar{t}}(\omega)- \Sigma^{scr}_{\bar{m}\bar{o}\bar{k};\bar{r}\bar{s}\bar{t}}\right]^{-1}_{mok;rst}
\label{Eqn:G3_3p}
\end{align}
%
%\begin{align}
%G_{mok;rst}^{\text{3p}}(\omega) &= \big[(\omega-\epsilon^{0}_{\bar{m}}-(\epsilon^{0}_{\bar{k}}-\epsilon^{0}_{\bar{o}})) \delta_{\bar{m}\bar{r}}\delta_{\bar{o}\bar{s}}\delta_{\bar{k}\bar{t}}
%\nonumber \\ & - (f_{\bar{m}}-f_{\bar{k}}) (f_{\bar{o}}-f_{\bar{k}}) \Sigma^{scr}_{\bar{m}\bar{o}\bar{k};\bar{r}\bar{s}\bar{t}}\big]_{mok;rst}^{-1}\nonumber\\
% &\times (f_r-f_t)(f_s-f_t).
%\end{align}
%
The expression in Eq.~\ref{Eqn:downfold} shows that:
1) although the multichannel self-energy is static the corresponding 1-body self-energy is dynamical thanks to $G^{\text{3p}}(\omega)$;
2) since $\Sigma^{1p/3p} = [\Sigma^{3p/1p}]^{\dagger}$, the screened $(3,1)$-MCDE is guaranteed to yield a positive-definite spectral function~\cite{Ste14};
3) although the multichannel self-energy $\Sigma_3$ only contains contributions that are of first order in the (screened) interaction, because of the inverse matrix in Eq.~\ref{Eqn:G3_3p}, 
the corresponding 1-body self-energy contains many contributions that are of infinite order in the interaction.
The same observations hold for the standard $(3,1)$-MCDE.
Finally, we note that other approximations to the one-body self-energy, such as the $GW$ and T-matrix self-energies, can be rewritten in the form given in Eq.~\ref{Eqn:downfold}.~\cite{Rom12,Berkelbach_JCP2021,Loo22}
%Therefore, by upfolding the one-body self-energy, a multichannel self-energy can be obtained for these approximations.
One can thus show that the $GW$ and T-matrix approximations only contain a small subset of the contributions contained in the (3,1)-SMCDE multichannel self-energy.

Because the multichannel self-energy is static, the screened MCDE (SMCDE) can be rewritten as an eigenvalue equation of the following effective Hamiltonian~\cite{riva_prl,riva_prb}
\begin{align}
    H^{\text{SMCDE}}=\begin{pmatrix}
        H^{1 \text p} & H^{\text 1p/3p} \\
        H^{\text 3p/1p} & H^{3 \text p, scr}
    \end{pmatrix},
\end{align}
where
\begin{align}
    H^{1 \text p}_{i;m}&=\epsilon_i\delta_{im},\\
    H^{\text 1p/3p}_{i;m>ok} &= (f_m - f_k)(f_o - f_k)\Sigma^{\text 1p/3p}_{i;mok} \\
     H^{\text 3p/1p}_{i>jl;m}&=   (f_i - f_l)(f_j - f_l) \Sigma^{\text 3p/1p}_{ijl;m} \\
    H^{3\text p, scr}_{i>jl;m>ok}&=(\epsilon^0_i-(\epsilon^0_l-\epsilon^0_j))\delta_{im}\delta_{jo}\delta_{lk} 
    \nonumber \\&+(f_i - f_l)(f_j - f_l)\Sigma^{3 \text p}_{ijl;mok}.
\end{align}
The spectral function $A(\omega)$ can then be obtained from the eigenenergies $E_{\lambda}$ and eigenfunctions $A_{\lambda}$ of $H^{SMCDE}$ according to
\begin{equation}
A(\omega) = \frac{1}{\pi}\sum_{i \in \text{1p}}
\left|
\text{Im} \sum_{\lambda} \frac{A^i_{\lambda} A^{*i}_{\lambda}}{\omega - E_{\lambda}+ i\eta\text{sgn}(E_{\lambda}-\mu)}
\right|
\end{equation}
where $\mu$ is the chemical potential, $\eta$ is a positive infinitesimal and $i \in \text{1p}$ means the sum is restricted to the 1-particle channel.

In practice we obtained the Hartree-Fock eigenenergies from a perturbative calculation on top of a density-functional calculation within the local-density approximation (LDA) 
and the matrix elements of the bare and screened Coulomb potentials have been calculated using the LDA orbitals.
The dielectric function required to evaluate the screened potential has been calculated within the random-phase approximation.
Finally, the predominant eigenvalues and eigenfunctions of the (3,1)-SMCDE Hamiltonian have been obtained using the iterative Lanczos-Haydock method~\cite{Hay72}.
For comparison we also calculated $GW$ spectral functions. We used a standard perturbative approach on top of an LDA calculation.

\begin{figure}[t]
    \centering
    \includegraphics[width=1.0\linewidth]{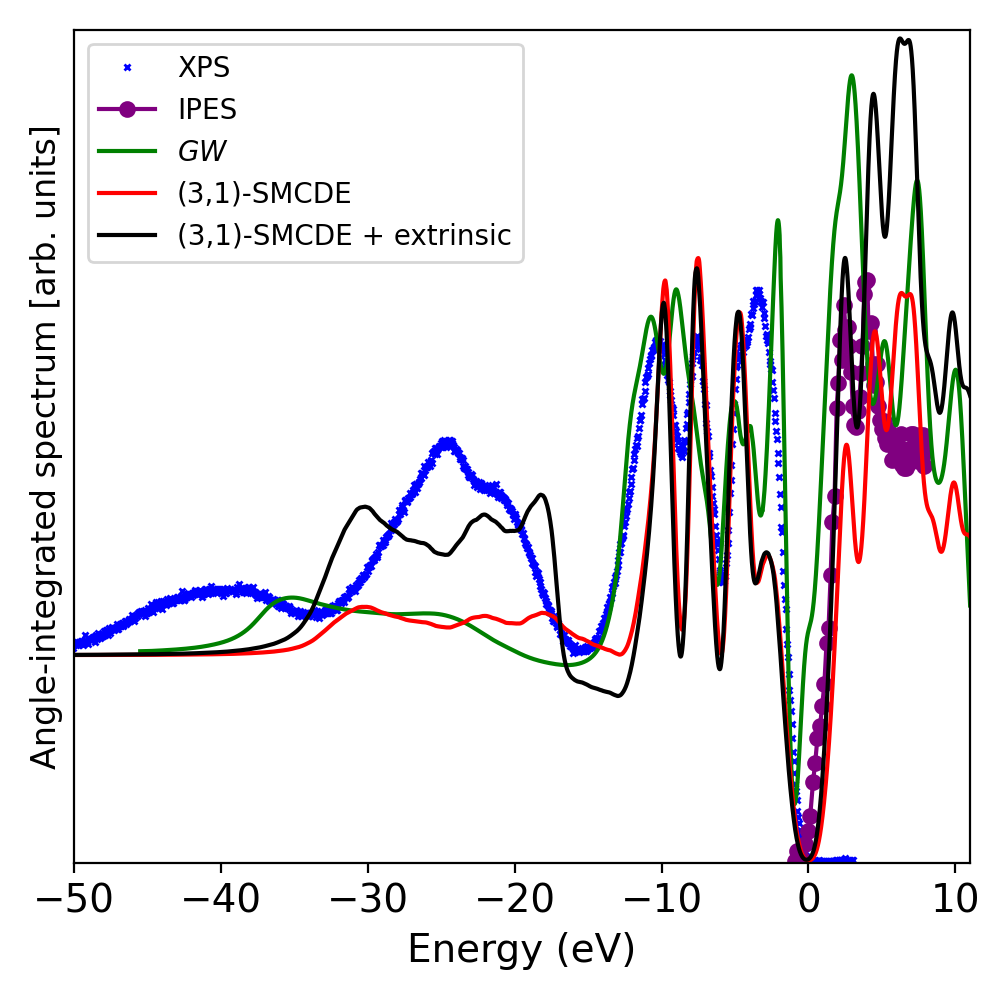}
    \caption{Comparison between theoretical calculations and experimental measurements of the angle-integrated spectrum of bulk silicon. The experimental data were taken from Ref.~\cite{Guz11} (XPS) and \cite{Straub_1985} (IPES).}
    \label{Fig1}
\end{figure}
\begin{figure}[h]
    \centering
    \includegraphics[width=1.0\linewidth]{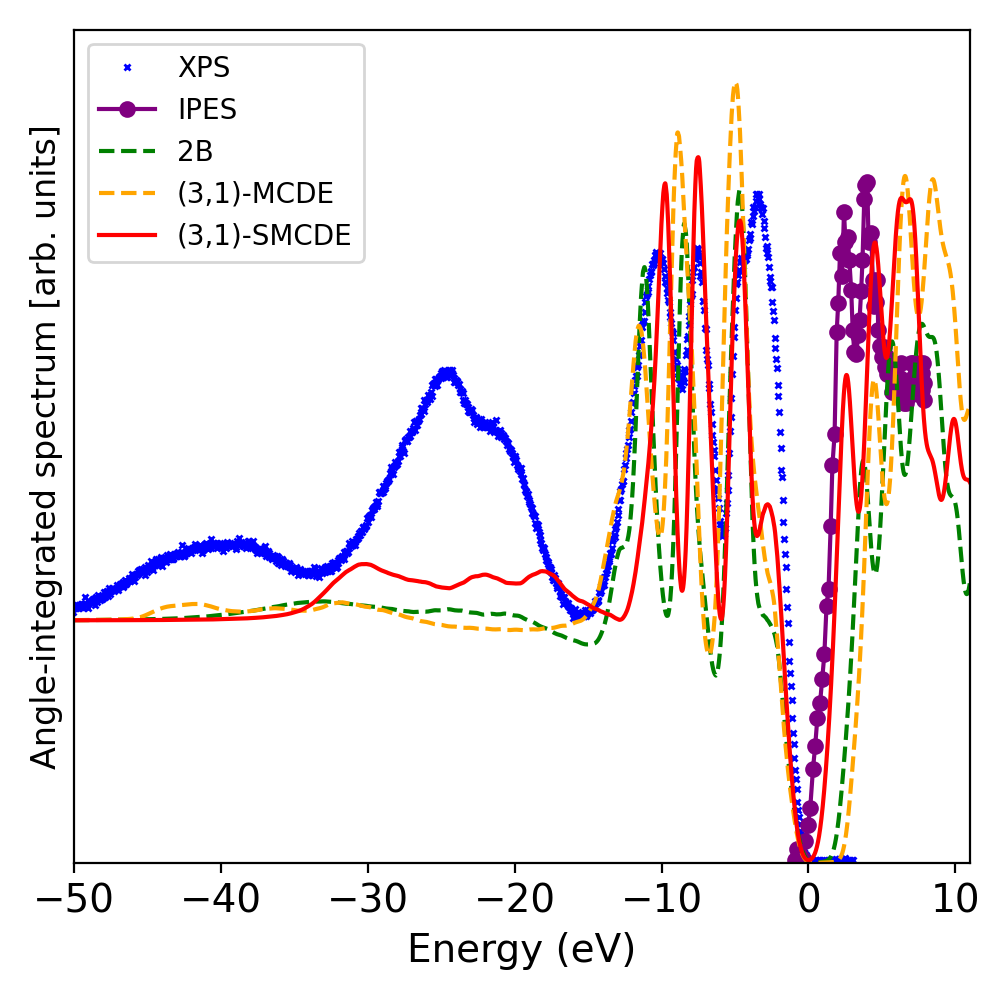}
    \caption{Comparison between theoretical calculations and experimental measurements of the angle-integrated spectrum of bulk silicon. The experimental data were taken from Ref.~\cite{Guz11} (XPS) and \cite{Straub_1985} (IPES).}
    \label{Fig2}
\end{figure}

In Fig.~{\ref{Fig1}} we compare the (3,1)-SMCDE spectral function of bulk silicon to the experimental direct and inverse photoemission spectra and to the $GW$ spectral function.
All spectra have been aligned at the valence-band maximum which has been set to zero.
We have included the photo-absorption cross sections in the theoretical spectra using the tabulated data of Ref.~\cite{Yeh_1985}.
In order to facilitate the comparison with experiment we have applied a Gaussian broadening of 0.4 eV to the theoretical spectra to account for the finite k-point sampling and we added a background to the theoretical spectra 
using the approach described in Ref.~\cite{Shirley_1972} to account for the secondary electrons in the direct photoemission experiment.
Finally, we also report a (3,1)-SMCDE spectrum that includes contributions due to the extrinsic losses and interference effects using a similar approach to that described in Refs.~\cite{Guz11,Lischner_2013}.
As with the approach in Ref.~\cite{Lischner_2013}, our approach accounts for the shift in spectral weight from the quasiparticles to the plasmon satellite but neglects changes in the shape of spectrum.

We see that the SMCDE spectral function reproduces the main features of the experimental spectra.
In particular, the position of the plasmon satellite centred around -25 eV  is correctly predicted by the (3,1)-SMCDE, unlike $GW$ which predicts it to be at much lower energy.
The positions of the quasiparticle energies both above and below the Fermi level as well as the band gap are also well-described by the (3,1)-SMCDE.
In particular, in the unoccupied part of the spectrum, the (3,1)-SMCDE captures the two peaks in the experimental spectrum around 2.4 and 4.2 eV~\cite{Straub_1985}.
Finally, we note that the (3,1)-SMCDE misses the small peak around -42 eV present in the experimental spectrum.
This peak is a replica of the satellite peak around -25 eV and cannot be described by the (3,1)-SMCDE as it describes 1$h$-1$e$ excitations but not 2$h$-2$e$ double excitations upon the removal of an electron.
In order to capture this replica we would have to use the (5,1)-SMCDE which also includes the 3$h$-2$e$ and 3$e$-2$h$ channels of the five-body Green's function since the 3$h$-2$e$ channel describes double excitations upon removing an electron.
Alternatively, one could use a frequency-dependent multichannel self-energy in the (3,1)-SMCDE.

To demonstrate the importance of introducing the screened potential in the MCDE, in Fig.~\ref{Fig2} 
we compare the spectra obtained with the (3,1)-SMCDE to those obtained with the (3,1)-MCDE as well as to spectra obtained within the second Born (2B) approximation.
The latter spectra can be obtained from our (3,1)-MCDE formulation by setting $\Sigma^{3\text p} = 0$. 
The (3,1)-MCDE equations then become equal to those obtained within second-order algebraic diagrammatic construction (ADC(2)) ~\cite{Sch83,Nie84} which is equivalent to 2B (also called GF2 in the literature).
We see that the quasiparticle features of the spectra are similar but the plasmon satellites in the (3,1)-MCDE and 2B spectra are centred at too low energy similar to the plasmon satellite in the $GW$ spectrum.

In conclusion, we have presented the screened multichannel Dyson equation to simulate both direct and inverse photoemission spectra.
By construction all interactions are screened in this equation, either directly or indirectly.
We showed that it correctly describes the main features of the direct and inverse photoemission spectra of bulk silicon.
Moreover, we showed that the proper screening of all interactions is crucial to correctly describe the position of the silicon plasmon satellite.
Therefore, when also considering the results we previously obtained~\cite{riva_prl,Paggi_JCP_2025}, the screened multichannel Dyson equation is a promising approach for the description of photoemission spectra of both weakly and strongly correlated systems.

\section*{Acknowledgment}
We thank the French Agence Nationale de la Recherche (ANR) for financial support (Grant Agreement ANR-22-CE29-0001 and ANR-22-CE30-0027).
JAB thanks Francesco Sottile for fruitful discussions. We thank Matteo Gatti for providing us with the experimental XPS data.
\end{document}